\documentclass[runningheads]{llncs}

\usepackage{booktabs} 
\usepackage{times}
\usepackage{latexsym}
\usepackage{booktabs}
\usepackage{amssymb}
\usepackage{dsfont}
\usepackage{caption}
\usepackage{multirow}
\usepackage{subfig}
\usepackage{amsmath}
\usepackage{todonotes}
\usepackage{csquotes}
\usepackage{hyperref}
\usepackage{url}
\usepackage{graphicx,wrapfig,lipsum}
\usepackage{soul}

\renewcommand{\theenumii}{\theenumi.\arabic{enumii}.}
\usepackage{arydshln}
\usepackage[markup=underlined]{changes}

\usepackage{xcolor,colortbl}
\usepackage{array}
\newcolumntype{L}[1]{>{\raggedright\let\newline\\\arraybackslash\hspace{0pt}}m{#1}}
\newcolumntype{C}[1]{>{\centering\let\newline\\\arraybackslash\hspace{0pt}}m{#1}}
\newcolumntype{R}[1]{>{\raggedleft\let\newline\\\arraybackslash\hspace{0pt}}m{#1}}

\begin{document}

\title{Neural-IR-Explorer: A Content-Focused Tool to~Explore~Neural~Re-Ranking~Results}

\author{Sebastian Hofst{\"a}tter \and Markus Zlabinger \and Allan Hanbury}

\authorrunning{S. Hofst{\"a}tter et al.}
%
\institute{TU Wien, Vienna, Austria \\ \email{\{\textit{first.last}\}@tuwien.ac.at}}

\maketitle

\begin{abstract}
In this paper we look beyond metrics-based evaluation of Information Retrieval systems, to explore the reasons behind ranking results. We present the content-focused Neural-IR-Explorer, which empowers users to browse through retrieval results and inspect the inner workings and fine-grained results of neural re-ranking models. The explorer includes a categorized overview of the available queries, as well as an individual query result view with various options to highlight semantic connections between query-document pairs. 

The Neural-IR-Explorer is available at:

{\small{\textit{\url{https://neural-ir-explorer.ec.tuwien.ac.at/}}}}

\end{abstract}

\section{Introduction}
\vspace{-0.3cm}

The prevalent evaluation of Information Retrieval systems, based on metrics that are averaged across a set of queries, distills a large variety of information into a single number. This approach makes it possible to compare models and configurations, however it also decouples the explanation from the evaluation. With the adoption of neural re-ranking models, where the scoring process is arguably more complex than traditional retrieval methods, the divide between result score and the reasoning behind it becomes even stronger. Because neural models learn based on data, they are more likely to evade our intuition about how their components should behave. Having a thorough understanding of neural re-ranking models is important for anybody who wants to analyze or deploy these models \cite{Hofstaetter2019_osirrc,Hofstaetter2019_sigir}

In this paper we present the Neural-IR-Explorer: a system to explore the output of neural re-ranking models. The explorer complements metrics based evaluation, by focusing on the content of queries and documents, and how the neural models relate them to each other. We enable users to efficiently browse the output of a batched retrieval run. We start with an overview page showing all evaluated queries. We cluster the queries using their term representations taken from the neural model. Users can explore each query result in more detail: We show the internal partial scores and content of the returned documents with different highlighting modes to surface the inner workings of a neural re-ranking model. Here, users can also select different query terms to individually highlight their connections to the terms in each document.

In our demo we focus on the kernel-pooling models KNRM \cite{Xiong2017} and TK \cite{Hofstaetter2019_trec} evaluated on the MSMARCO-Passage \cite{msmarco16} collection. The kernel-pooling makes it easy to analyze temporary scoring results. Finally, we discuss some of the insights we gained about the KNRM model using the Neural-IR-Explorer. The Neural-IR-Explorer is available at: {\small{\textit{\url{https://neural-ir-explorer.ec.tuwien.ac.at/}}}}.


\section{Related Work}
\vspace{-0.3cm}

Our work sits at the intersection of visual IR evaluation and the interpretability of neural networks with semantic word representations. The IR community mainly focused on tools to visualize result metrics over different configurations: \textit{CLAIRE} allows users to select and evaluate a broad range of different settings \cite{angelini2018claire}; \textit{AVIATOR} integrates basic metric visualization directly in the experimentation process \cite{giachelle2019progressive}; and the \textit{RETRIEVAL} tool provides a data-management platform for multimedia retrieval including differently scoped metric views \cite{ioannakis2017retrieval}. Lipani et al. \cite{lipani2017visual} created a tool to inspect different pooling strategies, including an overview of the relevant result positions of retrieval runs. 

From a visualization point of view term-by-term similarities are similar to attention, as both map a single value to a token. Lee et al. \cite{lee2017interactive} created a visualization system for attention in a translation task. Transformer-based models provide ample opportunity to visualize different aspects of the many attention layers used \cite{vig2019transformervis,coenen2019visualizing}. Visualizing simpler word embeddings is possible via a neighborhood of terms \cite{heimerl2018interactive}.


\begin{figure}[t]
   \includegraphics[
   width=\textwidth]{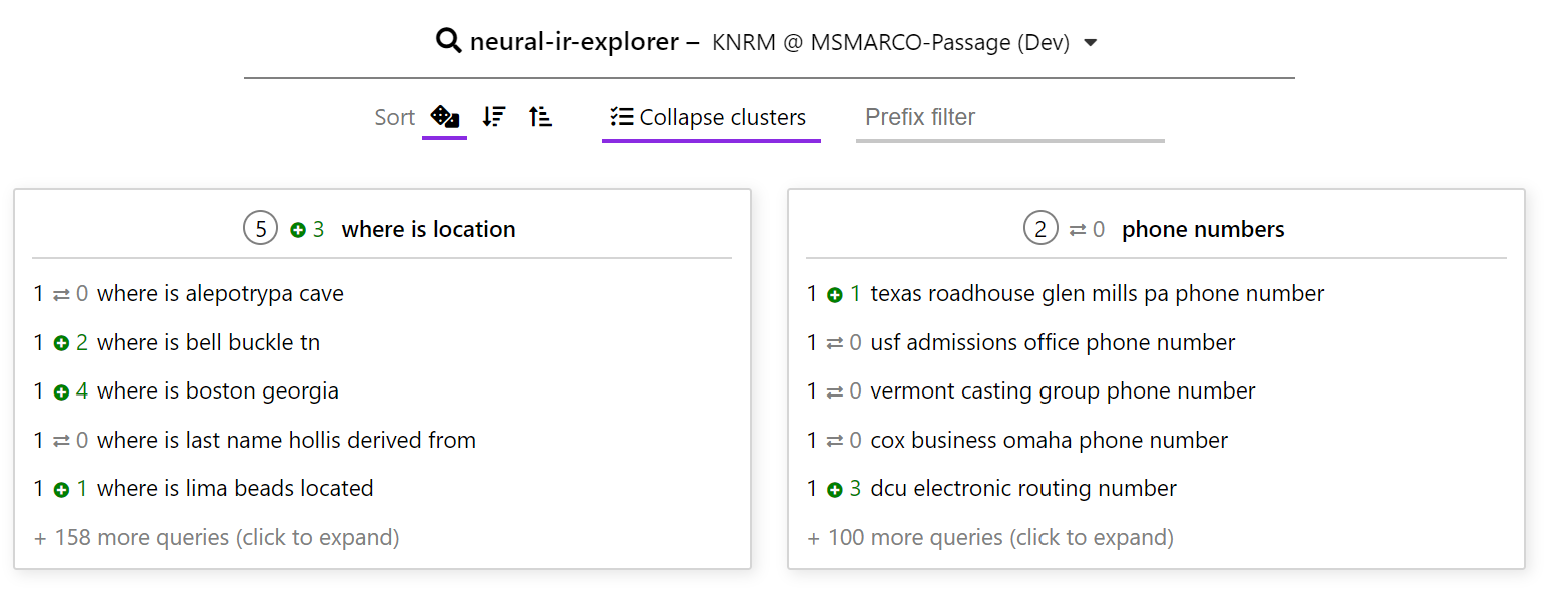}
    \centering
    \caption{Screenshot of the query cluster view with the controls at the top and two clusters}
    \label{fig:cluster_view}
    \vspace{-0.5cm}
\end{figure}

\vspace{-0.3cm}
\section{Neural-IR-Explorer}
\vspace{-0.3cm}
Now we showcase the capabilities of the Neural-IR-Explorer (Section \ref{sec:app}) and how we already used it to gain novel insights (Section \ref{sec:analysis}). The explorer displays data created by a batched evaluation run of a neural re-ranking model. The back-end is written in Python and uses Flask as web server; the front-end uses Vue.js. The source code is available at: {\small{\url{github.com/sebastian-hofstaetter/neural-ir-explorer}}}

\vspace{-0.3cm}
\subsection{Application}
\label{sec:app}
\vspace{-0.2cm}

When users first visit our website they are greeted with a short introduction to neural re-ranking and the selected neural model. We provide short explanations throughout the application, so that that new users can effectively use our tool. We expect this tool's audience to be not only neural re-ranking experts, but anyone who is interested in IR.

The central hub of the Neural-IR-Explorer is the query overview (Fig. \ref{fig:cluster_view}). We organize the queries by clustering them in visually separated cards. We collapse the cards to only show a couple of queries per default. This is especially useful for collections with a large number of queries, such as the MSMARCO collection we use in this demo (the DEV set contains over 6.000 queries). In the cluster header we display a manually assigned summary title, the median result of the queries, and median difference to the initial BM25 ranking, as this is the basis for the re-ranking. Each query is displayed with the rank of the first relevant document, the difference to BM25, and the query text. The controls at the top allow to sort the queries and clusters -- including a random option to discover new queries. Users can expand all clusters or apply a term-prefix filter to search for specific words in the queries. 

\begin{figure}[t]
   \includegraphics[trim={0cm 0cm 0cm 0cm} ,clip=True,
   width=\textwidth]{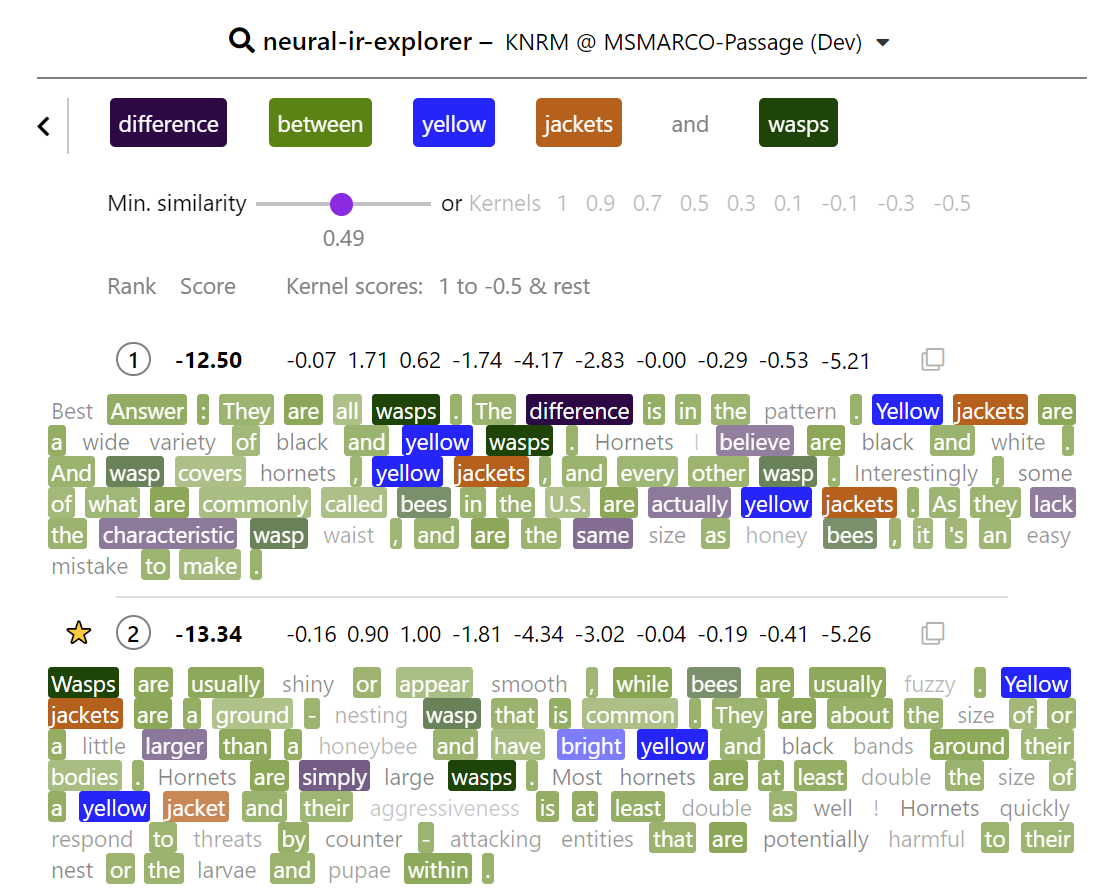}
    \centering
    \caption{Screenshot of the detailed query result view (with min. similarity highlighting)}
    \label{fig:query_view}
    \vspace{-0.5cm}
\end{figure}

Once a user clicks on a query, they are redirected to the query result view (Fig. \ref{fig:query_view}). Here, we offer an information rich view of the top documents returned by the neural re-ranking model. Each document is displayed in full with its rank, overall and kernel-specific scores. The header controls allow to highlight the connections between the query and document terms in two different ways. First, users can choose a minimum cosine similarity that a term pair must exceed to be colored, which is a simple way of exploring the semantic similarity of the word representations. Secondly, for kernel-pooling models that we support, we offer a highlight mode much closer to how the neural model sees the document: based on the association of a term to a kernel. Users can select one or more kernels and terms are highlighted based on their value after the kernel transformation.  

\begin{figure}[t]
   \includegraphics[
   width=\textwidth]{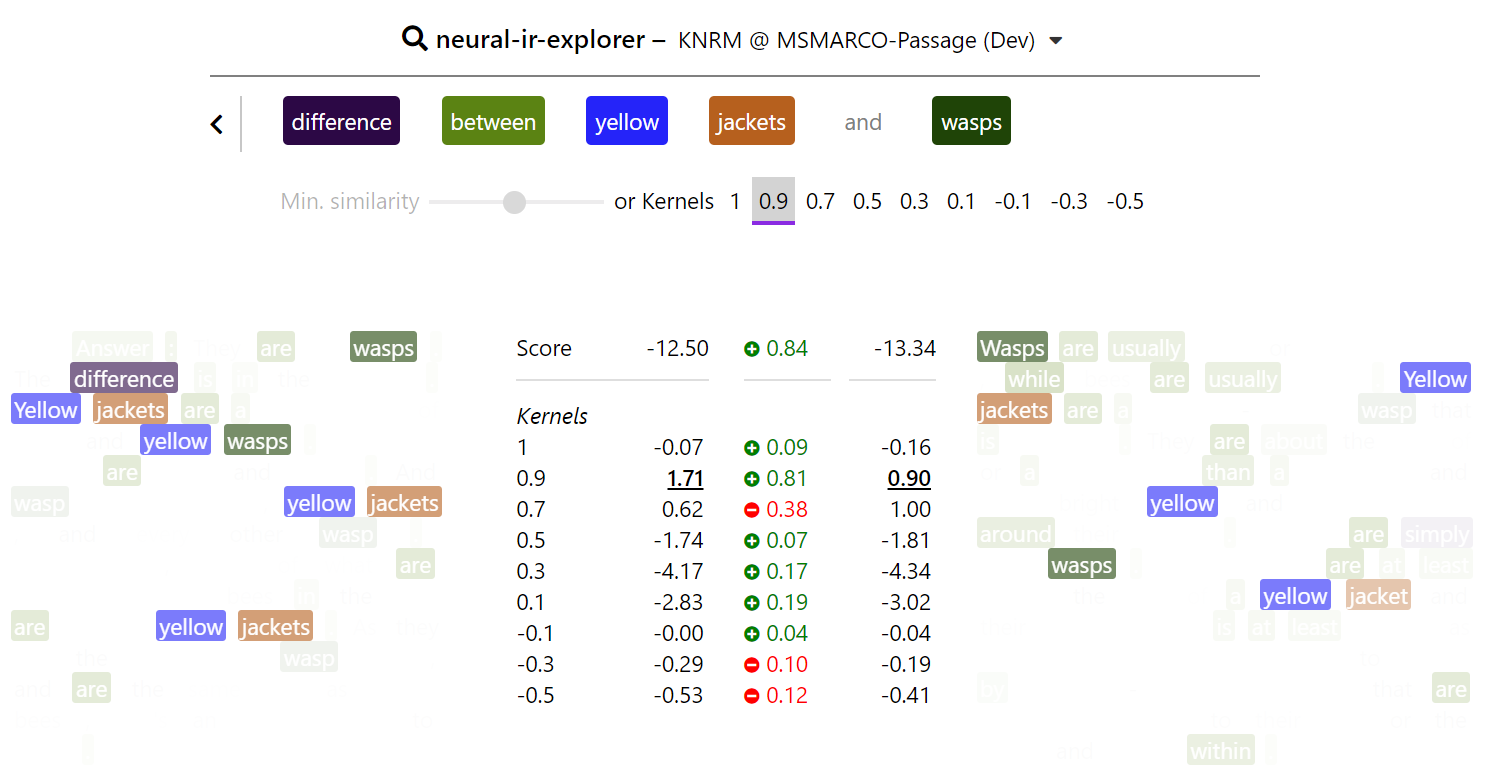}
    \centering
    \caption{Screenshot of the detailed side-by-side view of two documents}
    \label{fig:side2_view}
    \vspace{-0.5cm}
\end{figure}

Additionally, we enable users to select two documents and compare them side-by-side (Fig. \ref{fig:side2_view}). Users can highlight query-document connections as in the list view. Additionally, we display the different kernel-scores in the middle, so that users can effectively investigate which kernels of the neural model have the deciding influence of the different scores for the two documents. 

\vspace{-0.4cm}
\subsection{Neural Re-Ranking Model Analysis}
\label{sec:analysis}
\vspace{-0.2cm}

We already found the Neural-IR-Explorer to be a useful tool to analyze the KNRM neural model and understand its behaviors better. The KNRM model includes a kernel for exact matches (cosine similarity of exactly 1), however judging from the displayed kernel scores this kernel is not a deciding factor. Most of the time the kernels for 0.9 \& 0.7 (meaning quite close cosine similarities) are in fact the deciding factor for the overall score of the model. We assume this is due to the fact, that every candidate document (retrieved via exact matched BM25) contains exact matches and therefore it is not a differentiating factor anymore -- a specific property of the re-ranking task.



Additionally, the Neural-IR-Explorer also illuminates the pool bias \cite{lipani2016impact} of the MSMARCO ranking collection: The small number of judged documents per query makes the evaluation fragile. Users can see how relevant unjudged documents are actually ranked higher than the relevant judged documents, wrongly decreasing the model's score.

\vspace{-0.2cm}
\section{Conclusion}
\label{sec:conclusion}
\vspace{-0.2cm}

We presented the content-focused Neural-IR-Explorer to complement metric based evaluation of retrieval models. The key contribution of the Neural-IR-Explorer is to empower users to efficiently explore retrieval results in different depths. The explorer is a first step to open the black-boxes of neural re-ranking models, as it investigates neural network internals in the retrieval task setting. The seamless and instantly updated visualizations of the Neural-IR-Explorer offer a great foundation for future work inspirations, both for neural ranking models as well as how we evaluate them. 

\bibliographystyle{abbrv}
\bibliography{main} 

\end{document}